\documentstyle[epsfig,aps]{revtex}

\draft

\tightenlines

\draft

\begin{document}

\large

\title{Unnormalized nonextensive expectation value and zeroth law of thermodynamics}

\author{Qiuping A. Wang}
\address{Institut Sup\'{e}rieur des Mat\'{e}riaux du Mans, 44, Av.
Bartholdi, 72000 Le Mans, France \\ awang@ismans.univ-lemans.fr}

\maketitle

\begin{abstract}
We show an attempt to establish the zeroth law of thermodynamics
within the framework of nonextensive statistical mechanics based
on the classic normalization $\texttt{Tr}\hat{\rho}=1$ and the
unnormalized expectation $x=\texttt{Tr}\hat{\rho}^q\hat{x}$. The
first law of thermodynamics and the definition of heat and work
in this formalism are discussed.
\end{abstract}


\newpage

\section{Introduction}
A basic assumption of the equilibrium thermodynamics is the
existence of equilibrium state between two systems $A$ and $B$ at
which $T(A)=T(B)$ and $f(A)=f(B)$, where $T$ is the absolute
temperature and $f$ the generalized force (pressure, chemical
potential, electro-magnetic field, etc.), respectively. This is
the so called zeroth law of thermodynamics. Up to now, no
empirical evidence shows the contrary. So it is assumed that all
thermodynamic theories addressing systems at equilibrium should
conform with this law.

Maxwell-Boltzmann statistics ($MBS$) gives a beautiful
statistical interpretation of the zeroth law through Gibbs entropy
$S=-k\texttt{Tr}\hat{\rho}ln\hat{\rho}$ and Maxwell-Boltzmann
distribution $\hat{\rho}=\frac{1}{Z}e
^{-\beta(\hat{H}-\hat{f}_i\hat{x}_i)}$ for a given ensemble in the
representation of energy $\hat{H}$ and of the external variable
$\hat{x}_i$ ($\hat{x}_i$ can be volume, particle number, etc),
where $\beta=1/kT$ and $Z=\texttt{Tr}e
^{-\beta(\hat{H}-f_i\hat{x}_i)}$. Gibbs entropy can be recast as
$S=klnZ+k\beta E-k\beta f_ix_i$ ($E=\texttt{Tr}\hat{\rho}\hat{H}$
and $x_i=\texttt{Tr}\hat{\rho}\hat{x}_i$) where the term with
double index $f_ix_i$ signifies a summation over $i$. In this
formalism, a variation of the entropy of the total system $A+B$
is written as

\begin{eqnarray}                                        \label{a1}
dS(A+B) &=& \frac{\partial S(A)}{\partial E(A)}dE(A)+
\frac{\partial S(B)}{\partial E(B)}dE(B) + \frac{\partial
S(A)}{\partial x_i(A)}dx_i(A) + \frac{\partial S(B)}{\partial
x_i(B)}dx_i(B)\\ \nonumber
&=&k[\beta(A)dE(A)+\beta(B)dE(B)+\beta(A)f_i(A)dx_i(A)
+\beta(B)f_i(B)dx_i(B)]\\ \nonumber &=& 0
\end{eqnarray}
because we suppose $S(A+B)=S(A)+S(B)$ and $dS(A+B)=0$ at
equilibrium in the total system. Considering that
$dE(A+B)=d[E(A)+E(B)]=dE(A)+dE(B)=0$ and
$dx_i(A+B)=d[x_i(A)+x_i(B)]=dx_i(A)+dx_i(B)=0$, we get
$\beta(A)=\beta(B)$ and $f_i(A)=f_i(B)$.

This empirical law was believed\cite{Raggio} to be absent within
the nonextensive statistical mechanics ($NSM$) proposed by
Tsallis and co-workers \cite{Tsal1,Tsal2,Tsal3,Penni}. Recently,
some authors show that it can be established within $NSM$ by,
respectively, the approach with the standard normalization
$\texttt{Tr}\hat{\rho}=1$ and the normalized expectation
$x=\texttt{Tr}\hat{\rho}^q\hat{x}/\texttt{Tr}\hat{\rho}^q$ by
neglecting the nonextensive correlation term in $\hat{H}$
(assuming $E(A+B)=E(A)+E(B)$)\cite{Abe99,Abe01,Mart01,Mart00}, and
the approach of incomplete normalization
$\texttt{Tr}\hat{\rho}^q=1$ with the normalized expectation
$x=\texttt{Tr}\hat{\rho}^q\hat{x}$ in keeping the nonextensivity
in energy (i.e. $E(A+B)=E(A)+E(B)-(1-q)\beta E(A)E(B)$).

The present letter shows that the zero law can hold in the
formalism based on the normalization $\texttt{Tr}\hat{\rho}=1$
and the $unnormalized$ expectation
$x=\texttt{Tr}\hat{\rho}^q\hat{x}$. This formalism was proposed
by Tsallis and co-workers\cite{Tsal1,Tsal2} and had great success
in many applications\cite{Tsal4}. It also has the advantage to
give the simplest Legendre transformation. But recently,
scientists pay less attention to it due to its peculiar
properties such as, among others,
$\texttt{Tr}\hat{\rho}^q\hat{1}\neq 1$ and $E(A+B)\neq E(A)+E(B)$
for two independent systems having
$\hat{H}(A+B)=\hat{H}(A)+\hat{H}(B)$\cite{Tsal3}. These
peculiarities make one think that the zeroth law and the first law
may be disturbed in this formalism. In the following, we want to
show that the zero law can hold, despite the above inequality, in
the classic fashion. And the two basic processes of energy
change, heat and work, can be interpreted in the same way as in
$MBS$ analogy.

\section{The zeroth law}
The maximization of Tsallis entropy\footnote{Tsallis entropy is
given by
$S=-k\frac{\texttt{Tr}\hat{\rho}-\texttt{Tr}\hat{\rho}^q}{1-q} ,
(q \in R)$\cite{Tsal1}} subject to two constraints
$\alpha(\texttt{Tr}\hat{\rho}-1)$ and
$\beta(\texttt{Tr}\hat{\rho}^q\hat{H}-E)$ gives rise to the
following distribution function for canonical ensemble\cite{Tsal2}
\begin{eqnarray}                                        \label{a2}
\hat{\rho}=\frac{1}{Z}[1-(1-q)\beta\hat{H}]^{\frac{1}{1-q}}
\end{eqnarray}
where
\begin{eqnarray}                                        \label{a3}
Z=\texttt{Tr}[1-(1-q)\beta\hat{H}]^{\frac{1}{1-q}}.
\end{eqnarray}
where the Lagrange multiplier $k\beta=\frac{\partial S}{\partial
E}$. Supposing $\hat{\rho}(A+B)=\hat{\rho}(A)\hat{\rho}(B)$ for a
total system composed of two correlated subsystems $A$ and $B$
with the same $q$, we obtain,
\begin{eqnarray}                                        \label{a4}
S(A+B)=S(A)+ S(B)+\frac{1-q}{k}S(A)S(B)
\end{eqnarray}
and
\begin{eqnarray}                                        \label{a5}
E(A+B)=E(A)\texttt{Tr}\hat{\rho}^q(B)+
E(B)\texttt{Tr}\hat{\rho}^q(A)-(1-q)\beta E(A)E(B).
\end{eqnarray}
The relations Eq.(\ref{a4}) and (\ref{a5}) are to be considered as
two basic assumptions of the theory. We will establish the zero
law on this basis. From Eq.(\ref{a4}), we can write, for a small
variation of the total entropy :
\begin{eqnarray}                                        \label{a6}
dS(A+B) & = & [1+\frac{1-q}{k}S(B)]dS(A)+
[1+\frac{1-q}{k}S(A)]dS(B) \\ \nonumber
    & = & [1+\frac{1-q}{k}S(B)]
\frac{\partial S(A)}{\partial E(A)} dE(A) + [1+\frac{1-q}{k}S(A)]
\frac{\partial S(B)}{\partial E(B)} dE(B).
\end{eqnarray}
Because $dS(A+B)=0$, we get
\begin{eqnarray}                                        \label{a7}
\texttt{Tr}\hat{\rho}^q(B)\frac{\partial S(A)}{\partial E(A)}
dE(A) + \texttt{Tr}\hat{\rho}^q(A)\frac{\partial S(B)}{\partial
E(B)} dE(B)=0.
\end{eqnarray}
Now from Eq.(\ref{a2}), we easily verify that
\begin{eqnarray}                                        \label{a8}
\texttt{Tr}\hat{\rho}^q=Z^{1-q}+(1-q)\beta E.
\end{eqnarray}
So Eq.(\ref{a5}) becomes
\begin{eqnarray}                                        \label{a9}
E(A+B)=E(A)Z^{1-q}(B)+ E(B)Z^{1-q}(A)+(1-q)\beta E(A)E(B).
\end{eqnarray}
we can write
\begin{eqnarray}                                        \label{a10}
dE(A+B) &=& [Z^{1-q}(B)+(1-q)\beta E(B)]dE(A)\\
\nonumber &+& [Z^{1-q}(A)+(1-q)\beta E(A)]dE(B)
\end{eqnarray}
or, from Eq.(\ref{a8}),
\begin{eqnarray}                                        \label{a11}
dE(A+B) & = & \texttt{Tr}\hat{\rho}^q(B)dE(A)
+\texttt{Tr}\hat{\rho}^q(A)dE(B)=0
\end{eqnarray}
due to $dE(A+B)=0$ for the total system. Comparing Eq.(\ref{a11})
with Eq.(\ref{a7}), we get
\begin{eqnarray}                                        \label{a12}
\frac{\partial S(A)}{\partial E(A)} =\frac{\partial
S(B)}{\partial E(B)}
\end{eqnarray}
or $\beta(A)=\beta(B)$. The zeroth law of thermodynamics holds.
We can naturally define, as in $MBS$, $\frac{\partial S}{\partial
E}=\frac{1}{T}$ and write $T(A)=T(B)$ at equilibrium.

\section{The first law : Heat and work}
The above discussion suggests that the first law of
thermodynamics can be written as before for canonical ensemble :
\begin{eqnarray}                                        \label{c1}
dE=TdS+f_idx_i.
\end{eqnarray}
From Eq.(\ref{a8}), we easily find Helmoholtz free energy $F$ :

\begin{eqnarray}                                        \label{c2}
F=E-TS=-\frac{1}{\beta}\frac{Z^{1-q}-1}{1-q}
\end{eqnarray}
and
\begin{eqnarray}                                        \label{c3}
dF=-SdT+f_idx_i.
\end{eqnarray}

It is known that one of the beautiful interpretation of
thermodynamics given by statistical mechanics is the
understanding of the two kinds of process by which the energy of
a system can be changed : heat and work. In $MBS$, transferred
heat is interpreted as the energy change related to the
probability or population variation
($\texttt{Tr}\hat{H}d\hat{\rho}$). The work performed by the
surroundings on the system is related to the variation of the
energy of each state of the system
($\texttt{Tr}\hat{\rho}d\hat{H}$). We will show that this
interpretation can hold in the formalism discussed here.

From the unnormalized expectation mentioned above, we can write
\begin{eqnarray}                                        \label{b1}
dE=\texttt{Tr}\hat{H}d\hat{\rho}^q+\texttt{Tr}\hat{\rho}^q
d\hat{H}.
\end{eqnarray}
We will show that the first term at the right hand side can be
identified to heat ($TdS$) and the second term to work
($f_idx_i$).

With the help of Eq.(\ref{a2}), the first term of Eq.(\ref{b1})
can be recast as
\begin{eqnarray}                                        \label{b2}
\texttt{Tr}\hat{H}d\hat{\rho}^q
&=&\texttt{Tr}\frac{1-Z^{1-q}\hat{\rho}^{1-q}}
{(1-q)\beta}d\hat{\rho}^q \\ \nonumber
&=&\frac{1}{(1-q)\beta}[\texttt{Tr}d\hat{\rho}^q-
Z^{1-q}\texttt{Tr}\hat{\rho}^{1-q}d\hat{\rho}^q] \\ \nonumber
&=&\frac{1}{(1-q)\beta}[d\texttt{Tr}\hat{\rho}^q-
Z^{1-q}q\texttt{Tr}d\hat{\rho}].
\end{eqnarray}
Considering $\texttt{Tr}d\hat{\rho}=0$, we get
\begin{eqnarray}                                        \label{b3}
\texttt{Tr}\hat{H}d\hat{\rho}^q
&=&\frac{1}{(1-q)\beta}d\texttt{Tr}\hat{\rho}^q \\ \nonumber
&=&\frac{1}{k\beta}d[-k\frac{1-\texttt{Tr}\hat{\rho}^q}{1-q}] \\
\nonumber &=& TdS \\ \nonumber &=& dQ,
\end{eqnarray}
where $dQ$ is the heat transferred from the surroundings to the
system.

New let us see the second term in Eq.(\ref{b1}). We can recast it
as
\begin{eqnarray}                                        \label{b4}
\texttt{Tr}\hat{\rho}^q d\hat{H}
&=&\texttt{Tr}\hat{\rho}^q\frac{\partial \hat{H}}{\partial
x_i}dx_i \\ \nonumber
&=&\frac{1}{Z^q}\texttt{Tr}[1-(1-q)\beta\hat{H}]^{\frac{q}{1-q}}
\frac{\partial \hat{H}}{\partial x_i}dx_i \\ \nonumber
&=&-\frac{1}{Z^q\beta}\texttt{Tr} \{ \frac{\partial}{\partial
x_i}[1-(1-q)\beta\hat{H}]^{\frac{1}{1-q}} \}_\beta dx_i \\
\nonumber &=&-\frac{1}{\beta}\frac{1}{Z^q} \{ \frac{\partial
Z}{\partial x_i} \}_\beta dx_i \\ \nonumber &=&-\frac{1}{\beta}
\{\frac{\partial}{\partial x_i}\frac{Z^{1-q}-1}{1-q} \}_\beta
dx_i.
\end{eqnarray}
Considering Eqs.(\ref{c2}) and (\ref{c3}), we obtain finally

\begin{eqnarray}                                        \label{b5}
\texttt{Tr}\hat{\rho}^q d\hat{H} &=& \{\frac{\partial F}{\partial
x_i} \}_\beta dx_i \\ \nonumber &=& f_idx_i \\ \nonumber &=& dW
\end{eqnarray}
where $dW$ is the work performed by the surroundings on the
system. So the statistical interpretation of heat and work
remains the same in $NSM$ as in $MBS$.

\section{Conclusion}
We shown that within the formalism of $NSM$ based on the
normalization $\texttt{Tr}\hat{\rho}=1$ and the unnormalized
expectation $x=\texttt{Tr}\hat{\rho}^q\hat{x}$, the zeroth law
can hold as in $MBS$. We need not generalized (or physical)
temperature, heat and
forces\cite{Wang00,Wang01,Abe99,Mart01,Mart00} different from
that in $MBS$ to keep this basic assumption of thermodynamics.
The temperature defined by $T=\frac{\partial E}{\partial S}$
remains the measure of thermodynamic equilibrium. Heat and work
are interpreted just as in $MBS$, which seems difficult in the
formalisms with normalized expectations. The first law remains the
same. This {\it unnormalized formalism} finally gives the
simplest nonextensive thermodynamic formalism which seems
reassuring. This tells us that it is perhaps worthy to evaluate
again the unnormalized expectation and to try to find what is
hidden behind the peculiarities which remain to be explained.


\end{document}